\documentclass{article}
\usepackage[preprint]{spconf}
\usepackage{amsmath,graphicx}
\usepackage{subfigure}
\usepackage{multirow}
\usepackage{boldline}

\toappear{978-1-6654-7189-3/22/\$31.00~\copyright~2023 IEEE\hfill}

\title{Exploring WavLM on Speech Enhancement}
\name{\parbox{0.6\linewidth}{\centering
Hyungchan Song$^{1,*}$, Sanyuan Chen$^2$, Zhuo Chen$^3$, Yu Wu$^2$,
Takuya Yoshioka$^3$, Min Tang$^3$, Jong Won Shin$^1$, Shujie Liu$^2$ \thanks{$*$ This work was done during internship.} \thanks{This work was partly supported by Institute for Information \& communications Technology Promotion (IITP) grant funded by the Korea government (MSIP) (No.2021-0-01696, High-Potential Individuals Global Training Program) and Microsoft Research Asia.}}}

\address{$^1$Gwanju Institute of Science and Technology, Republic of Korea\\
    $^2$Microsoft, China\\
  $^3$Microsoft, USA}

\begin{document}

\maketitle

\begin{abstract}
There is a surge in interest in self-supervised learning approaches for end-to-end speech encoding in recent years as they have achieved great success. Especially, WavLM showed state-of-the-art performance on various speech processing tasks. To better understand the efficacy of self-supervised learning models for speech enhancement, in this work, we design and conduct a series of experiments with three resource conditions by combining WavLM and two high-quality speech enhancement systems. Also, we propose a regression-based WavLM training objective and a noise-mixing data configuration to further boost the downstream enhancement performance. The experiments on the DNS challenge dataset and a simulation dataset show that the WavLM benefits the speech enhancement task in terms of both speech quality and speech recognition accuracy, especially for low fine-tuning resources. For the high fine-tuning resource condition, only the word error rate is substantially improved.
\end{abstract}
\begin{keywords}
self-supervised learning, speech enhancement, fine-tuning 
\end{keywords}
\section{Introduction}
In the field of natural language processing and computer vision, self-supervised learning (SSL) approaches have been proposed to learn universal useful representations, which benefit a variety of downstream tasks. Recently, SSL approaches for speech audio processing~\cite{schneider2019wav2vec,baevski2019vq,baevski2020wav2vec2,hsu2021hubert,wang2021unispeech,chung2021w2v,wang2021unispeechas} have been proposed, focusing on phoneme classification and automatic speech recognition (ASR). Especially, inspired by the masked language model~\cite{kenton2019bert}, the masked predictive SSL approaches have achieved great success in various speech processing downstream tasks. Unlike previous work that designs SSL and unsupervised learning approaches for certain tasks, e.g. speech enhancement~\cite{zezario2020self,qiu2021self,sivaraman2022efficient}, the wav2vec 2.0~\cite{baevski2020wav2vec2}, HuBERT~\cite{hsu2021hubert}, Unispeech-SAT~\cite{chen2022unispeech}, and WavLM~\cite{chen2022wavlm} models are task agnostic, which serve various downstream tasks with the same model.

Although WavLM showed state-of-the-art performance in Speech processing Universal PERformance Benchmark (SUPERB)~\cite{yang21c_interspeech}, its improvement on speech enhancement demonstrates a different trend from other tasks such as ASR.
Only 0.1 PESQ improvement is observed against the fbank baseline even when a large SSL model is integrated. This observation can also be found for other high-quality SSL models in SUPERB evaluation.  
Therefore, it is non-trivial to further understand the impact of task-agnostic pre-trained models for speech enhancement. 
Unlike classification-based tasks such as speech or speaker recognition, 
speech enhancement requires the model to estimate continuous denoised speech. 
As most state-of-the-art SSL pre-trained models employ a classification/prediction-based objective function, which potentially mismatches the continuous nature of the enhancement task, sub-optimum fine-tuning results might be obtained by a simple combination. Therefore, an SSL objective function that is more aligned with speech enhancement is also worth exploring.

\begin{figure*}[t]
    \centering
    \subfigure[]{
    \includegraphics[height=7cm]{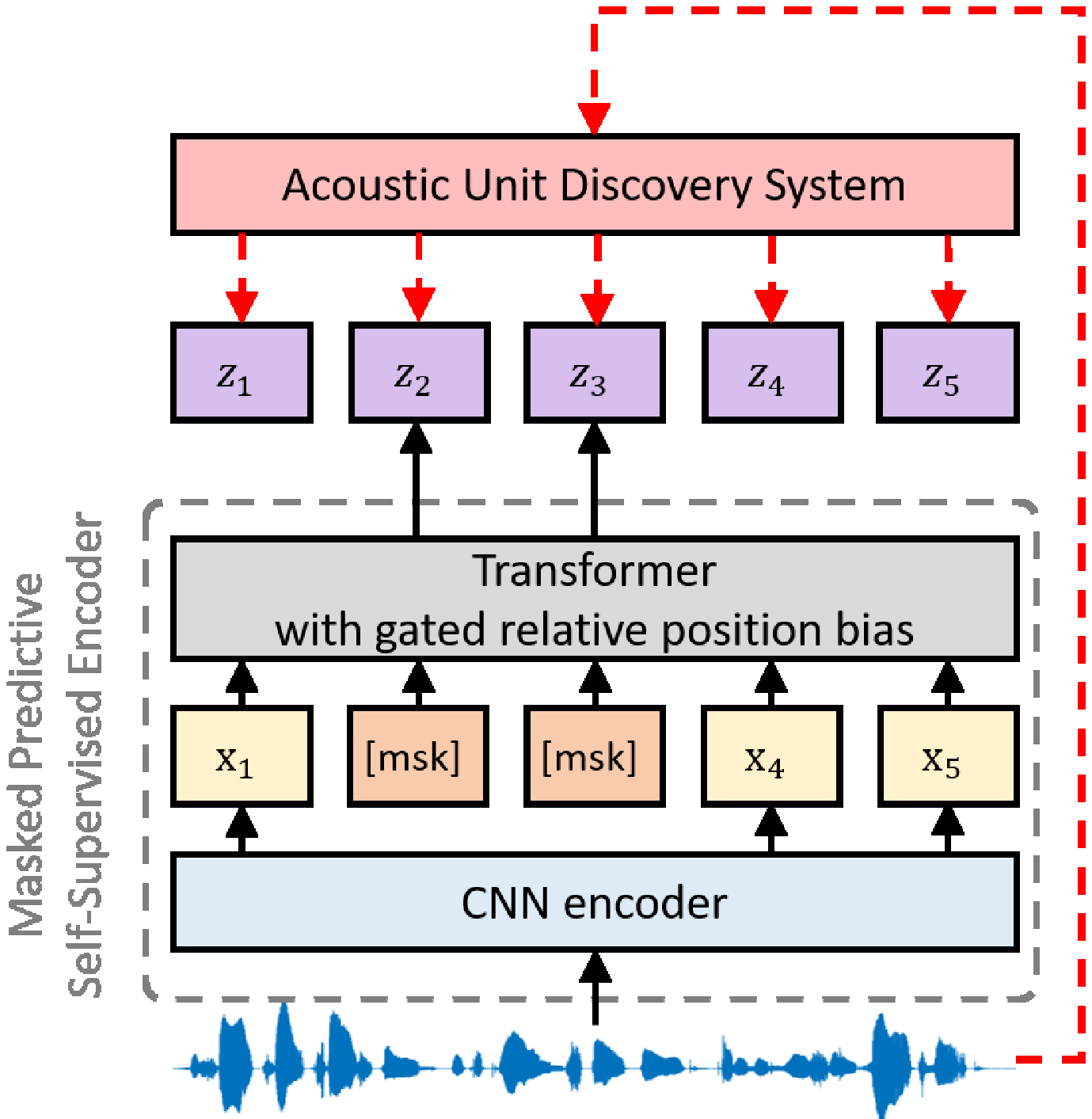}
    }\hspace{20mm}
    \subfigure[]{
    \includegraphics[height=7cm]{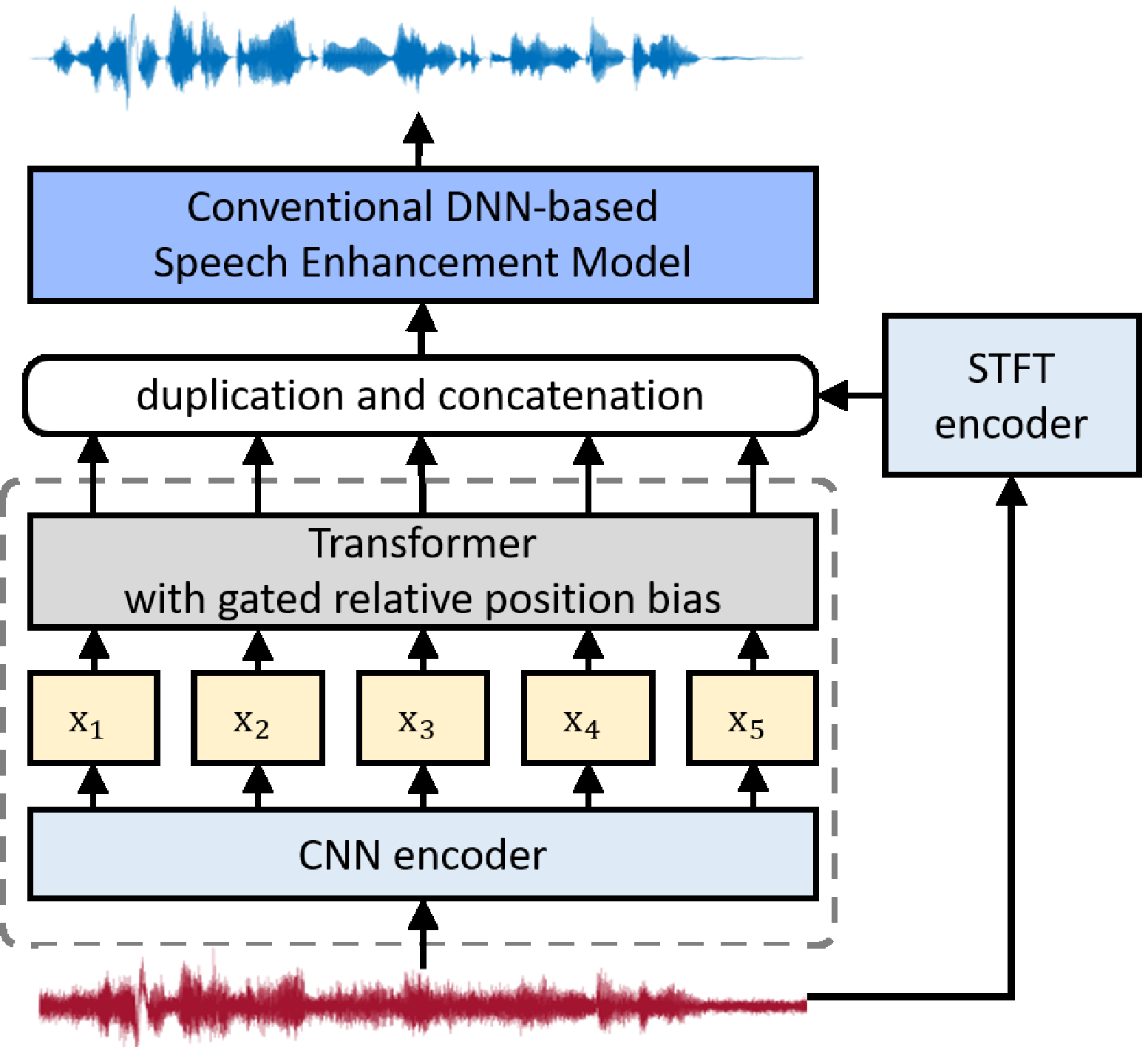}
    }
    \caption{An illustration of (a) the pre-training stage of WavLM and (b) the fine-tuning stage for speech enhancement. The masked predictive self-supervised encoder is identical in both stages.}
    \label{fig:illust}
\end{figure*}

To answer these questions, in this work, we design and conduct a series of experiments, by combining WavLM or its variants with two high-quality speech enhancement systems under different data conditions. Specifically, three scenarios are considered in this work: 1) low fine-tuning resource, 2) high fine-tuning resource, and 3) low pre-training and fine-tuning resource. 
Meanwhile, we propose a regression-based WavLM objective variant, where the network is optimized in an unsupervised fashion to predict the continuous output for the masked region from the input signal. A noise mixture training scheme is also explored, by randomly mixing additional noise clips to the input unlabeled speech during the SSL pre-training stage. 

In our evaluation with the DNS challenge dataset~\cite{reddy21_interspeech} and a simulation dataset, we found that the WavLM pre-trained model significantly improved the downstream speech enhancement in both speech quality and word error rate (WER) for all the low-resource scenarios.
For the high-resource scenarios, we observed a different trend for the WER and speech quality metrics, where the former still benefited from the SSL pre-trained model, while the latter only had minor improvements even with a large-scale WavLM. 
With the proposed regression-based WavLM model and noise mixing training strategy, better performance was observed for all conditions.  
Finally, we observed that the enhancement task was insensitive to the pre-training scale, where WavLM pre-trained with 960 hours and WavLM pre-trained with 94k hours showed similar performance.

\section{Method}

Based on the review of WavLM, in this section, we introduce the proposed regression loss and noise mixing training, followed by our exploration design in different conditions according to the amount of speech or noise.
Fig.~\ref{fig:illust} illustrates the WavLM pre-training stage and the speech enhancement fine-tuning stage.

\subsection{WavLM}

The WavLM~\cite{chen2022wavlm} model inspired by HuBERT~\cite{hsu2021hubert} contains two main networks as follows: a CNN encoder and a Transformer~\cite{vaswani2017attention} with $L$ blocks. During training, some frames of the CNN encoder output $\mathbf{x}$ are masked randomly and fed to the Transformer as input. The Transformer is optimized to predict the discrete target sequence $\mathbf{z}$, in which each $z_t \in [C]$ is a $C$-class categorical variable. The distribution over the classes is parameterized with
\begin{equation}
    p(c|\mathbf{h}_t)=\frac{\exp(\mathrm{sim}(\mathbf{W}^P\mathbf{h}^{L}_{t},\mathbf{e}_c)/\tau)}{\sum_{c'=1}^{C}\exp(\mathrm{sim}(\mathbf{W}^P\mathbf{h}^{L}_{t},\mathbf{e}_{c'})/\tau)}
\end{equation}
where $\mathbf{W}^P$ is a projection matrix, $\mathbf{h}^{L}_{t}$ is the output hidden state for step $t$, $\mathbf{e}_c$ is the embedding for class $c$, $\mathrm{sim}(a,b)$ means the cosine similarity between $a$ and $b$, and $\tau=0.1$ scales the logit. The prediction loss is applied over only masked regions, which forces the model to learn a combined acoustic and language model over the continuous inputs. In this work, the gated relative position bias~\cite{chi2022xlm} is employed to improve the performance of the Transformer, which is encoded based on the offset between the ``key'' and ``query'' in the self-attention modules of the Transformer. In this paper, we use WavLM Base+ of~\cite{chen2022wavlm} as the pre-trained model, which was trained on both non-mixed and mixed utterances with the probability of the latter being 0.1.

\subsection{Regression Loss and Noise Mixing Training}

The WavLM aims to predict short window phonetic units, which are well correlated with phoneme units, thus leading to a significant performance boost for classification-based downstream tasks. However, such a phonetic unit often ignores speech features such as pitch, tone, emotions, etc, which can be important for speech enhancement. To capture these features, we design a regression-based WavLM objective function. As with an enhancement fine-tuning network, the proposed regression WavLM loss function $\mathcal{L}_{reg}$ uses the L2 loss between clean 80-dimensional fbank $\textbf{z}_{fbank,t}$ and the projected latent representation $\textbf{W}^P\textbf{h}_t^L$ over the masked frames as follows:
\begin{equation}
    \mathcal{L}_{reg}(t) = \arg \min_{\textbf{W}^P\textbf{h}_t^L} (\textbf{W}^P\textbf{h}_t^L - \textbf{z}_{fbank,t})^2
\end{equation}

To further improve the pre-training generalization, we introduce a noise-mixing data configuration in the pre-training stage. 
Following~\cite{chen2022wavlm}, for each input speech utterance, we uniformly sample a noise clip from DNS's noise set, crop it to a random length, and then mix it with the input utterance at a random starting point with an energy ratio sampled from the uniform distribution $\mathcal{U}(-5, 20)$ dB.
Note that, as the data are unlabeled, the additional noise can also be applied to noisy input.

\subsection{Fine-tuning and Exploration Design}
In all experiments, a simulation dataset is used for fine-tuning, where the clean speech is sampled from the clean corpus of~\cite{braun2021towards}, while the DNS noise~\cite{reddy21_interspeech} (total 181 hours) is selected as the noise source.
In the fine-tuning stage, we combine pre-trained WavLM or its variants with an LSTM or Conformer enhancement model~\cite{chen2021continuous} as the task-specific network for speech enhancement. The LSTM model consists of a CNN layer and 3 bi-directional LSTM layers of 512 hidden units.
The Conformer architecture and setting are the same as conformer-based model of~\cite{chen2021continuous}. The input feature for fine-tuning is the concatenation of a noisy magnitude of STFT and a latent representation of the pre-training encoder for each frame, where the pre-training feature is processed by frame duplication as in~\cite{chen2022wavlm} to synchronize with the fine-tuning network.
The frequency magnitude mask is selected as the fine-tuning network output, with the L2 signal restoration loss~\cite{braun2021consolidated} as the objective.

It should be noted that, in this work, we don't use more advanced time-domain or complex-valued models~\cite{defossez2020real,kim2021se,hu2020dccrn,lv21_interspeech,li2021real,tzinis2022remixit} for fine-tuning, as the phase information is not considered in WavLM.
Adding phase modeling introduces additional variation, which is not the main focus of this work. 
To explore the potential and limitation of WavLM for speech enhancement, we designed three conditions according to the amount of speech or noise as described below. 

\subsubsection{Low fine-tuning resource}
We first design experiments to find out how much information can pre-trained WavLM models provide for scenarios with the limited fine-tuning resource. Two scenarios are considered for this condition: a limited speech scenario and a limited noise scenario. 
On the limited amount of speech resource setup, we restrict the speech resource to only 10 hours and use the full noise resource for fine-tuning. 
On the other hand, on the limited amount of noise resource setup, we limit the noise resource to a 10\% subset (18 hours) and use the full speech resource. 

\subsubsection{High fine-tuning resource}
The high fine-tuning resource setup uses a large-scale and high-quality simulated dataset for fine-tuning to evaluate the potential of WavLM on speech enhancement in a resource-rich setup. 
The dataset is described in~\cite{braun2021towards} and consists of around 1,000 hours of paired noisy and clean speech samples with various noise and room impulse response (RIR) conditions.

\subsubsection{Low pre-training and fine-tuning resource}
To examine the impact that the pre-training data quantity has on the enhancement performance, we limit the amount of both the pre-training and fine-tuning resources here.
In contrast to the huge amount of the pre-training data described in~\cite{chen2022wavlm}, we train WavLM with only LibriSpeech 960 hours in this setup to investigate the feasibility of the low pre-training resources for speech enhancement.
To see this, we compare this setup with the first low fine-tuning resource setup in Sec. 2.3.1.

\begin{table*}[t] \small
\centering
\caption{Results on the low fine-tuning resource setup with WavLM and its variants. The reg. and m.n denote the regression loss and noise mixing training strategy, respectively.}
\begin{tabular}{l|l|lll|lllll}
\hline \hline
\multicolumn{1}{c|}{\multirow{2}{*}{Model}} & \multicolumn{1}{c|}{\multirow{2}{*}{\begin{tabular}[c]{@{}c@{}}Pre-training \\ data\end{tabular}}} & \multicolumn{3}{c|}{DNS3 test data} & \multicolumn{5}{c}{Simulated test data} \\ \cline{3-10} 
& & \multicolumn{1}{c|}{SIG} & \multicolumn{1}{c|}{BAK} & \multicolumn{1}{c|}{OVR} & \multicolumn{1}{c|}{SIG} & \multicolumn{1}{c|}{BAK} & \multicolumn{1}{c|}{OVR} & \multicolumn{1}{c|}{SDR} & WER \\ \hline

Noisy   &  \multicolumn{1}{c|}{-} & \multicolumn{1}{c|}{3.776} & \multicolumn{1}{c|}{3.208} & \multicolumn{1}{c|}{3.114} &  \multicolumn{1}{c|}{3.837} & \multicolumn{1}{c|}{2.976} & \multicolumn{1}{c|}{3.098} & \multicolumn{1}{c|}{3.598} & \multicolumn{1}{r}{24.721}  \\ \hline
Oracle magnitude mask &\multicolumn{1}{c|}{-} & \multicolumn{1}{c|}{-} & \multicolumn{1}{c|}{-} & \multicolumn{1}{c|}{-} &  \multicolumn{1}{c|}{3.918} & \multicolumn{1}{c|}{4.151} & \multicolumn{1}{c|}{3.539} & \multicolumn{1}{c|}{12.537}    &  \multicolumn{1}{r}{2.596}   \\ \hline \hline
\multicolumn{9}{c}{\textbf{\textit{Limited amount of speech resource setup (10 hours)}}} \\ \hline \hline
LSTM        & \multicolumn{1}{c|}{-} & \multicolumn{1}{c|}{3.689} & \multicolumn{1}{c|}{3.687} & \multicolumn{1}{c|}{3.110} & \multicolumn{1}{c|}{3.708} & \multicolumn{1}{c|}{3.553} & \multicolumn{1}{c|}{3.105} & \multicolumn{1}{c|}{6.292}    & \multicolumn{1}{r}{27.446} \\ \hline
LSTM with WavLM        & \multicolumn{1}{c|}{94 kh} & \multicolumn{1}{c|}{3.747} & \multicolumn{1}{c|}{\textbf{3.858}} & \multicolumn{1}{c|}{3.217} & \multicolumn{1}{c|}{\textbf{3.795}} & \multicolumn{1}{c|}{3.733} & \multicolumn{1}{c|}{\textbf{3.227}} & \multicolumn{1}{c|}{7.402}    & \multicolumn{1}{r}{24.569} \\ \hline
LSTM with WavLM m.n.  & \multicolumn{1}{c|}{94 kh} & \multicolumn{1}{c|}{3.733} & \multicolumn{1}{c|}{3.817} & \multicolumn{1}{c|}{3.176} & \multicolumn{1}{c|}{3.769} & \multicolumn{1}{c|}{3.705} & \multicolumn{1}{c|}{3.180} & \multicolumn{1}{c|}{7.310}    & \multicolumn{1}{r}{24.048} \\ \hline
LSTM with WavLM reg.       & \multicolumn{1}{c|}{94 kh} & \multicolumn{1}{c|}{3.714} & \multicolumn{1}{c|}{3.811} & \multicolumn{1}{c|}{3.184} & \multicolumn{1}{c|}{3.774} & \multicolumn{1}{c|}{3.702} & \multicolumn{1}{c|}{3.177} & \multicolumn{1}{c|}{7.387}    & \multicolumn{1}{r}{23.633} \\ \hline
LSTM with WavLM reg. \& m.n.  & \multicolumn{1}{c|}{94 kh} & \multicolumn{1}{c|}{\textbf{3.750}} & \multicolumn{1}{c|}{3.853} & \multicolumn{1}{c|}{\textbf{3.219}} & \multicolumn{1}{c|}{3.793} & \multicolumn{1}{c|}{\textbf{3.750}} & \multicolumn{1}{c|}{3.225} & \multicolumn{1}{c|}{\textbf{7.697}}    & \multicolumn{1}{r}{\textbf{20.622}} \\ \hlineB{3}
LSTM with WavLM reg.       & \multicolumn{1}{c|}{960 h} & \multicolumn{1}{c|}{3.716} & \multicolumn{1}{c|}{3.758} & \multicolumn{1}{c|}{3.156} & \multicolumn{1}{c|}{3.746} & \multicolumn{1}{c|}{3.612} & \multicolumn{1}{c|}{3.156} & \multicolumn{1}{c|}{6.847}    & \multicolumn{1}{r}{25.931} \\ \hline
LSTM with WavLM reg. \& m.n. & \multicolumn{1}{c|}{960 h} & \multicolumn{1}{c|}{\textbf{3.753}} & \multicolumn{1}{c|}{\textbf{3.854}} & \multicolumn{1}{c|}{\textbf{3.220}} & \multicolumn{1}{c|}{\textbf{3.794}} & \multicolumn{1}{c|}{\textbf{3.760}} & \multicolumn{1}{c|}{\textbf{3.221}} & \multicolumn{1}{c|}{\textbf{7.777}}    & \multicolumn{1}{r}{\textbf{19.821}} \\ \hlineB{3}
Conformer  & \multicolumn{1}{c|}{-} & \multicolumn{1}{c|}{3.731} & \multicolumn{1}{c|}{3.753} & \multicolumn{1}{c|}{3.187} & \multicolumn{1}{c|}{3.780} & \multicolumn{1}{c|}{3.689} & \multicolumn{1}{c|}{3.212} & \multicolumn{1}{c|}{6.902}    & \multicolumn{1}{r}{27.018} \\ \hline
Conformer with WavLM   & \multicolumn{1}{c|}{94 kh} & \multicolumn{1}{c|}{3.729} & \multicolumn{1}{c|}{3.797} & \multicolumn{1}{c|}{3.199} & \multicolumn{1}{c|}{3.773} & \multicolumn{1}{c|}{3.703} & \multicolumn{1}{c|}{3.216} & \multicolumn{1}{c|}{7.260}    & \multicolumn{1}{r}{25.222} \\ \hline
Conformer with WavLM m.n  & \multicolumn{1}{c|}{94 kh} & \multicolumn{1}{c|}{3.745} & \multicolumn{1}{c|}{3.813} & \multicolumn{1}{c|}{3.213} & \multicolumn{1}{c|}{\textbf{3.812}} & \multicolumn{1}{c|}{3.751} & \multicolumn{1}{c|}{\textbf{3.252}} & \multicolumn{1}{c|}{7.592}    & \multicolumn{1}{r}{24.204} \\ \hline
Conformer with WavLM reg.  & \multicolumn{1}{c|}{94 kh} & \multicolumn{1}{c|}{\textbf{3.746}} & \multicolumn{1}{c|}{3.821} & \multicolumn{1}{c|}{3.194} & \multicolumn{1}{c|}{3.775} & \multicolumn{1}{c|}{3.713} & \multicolumn{1}{c|}{3.220} &  \multicolumn{1}{c|}{7.368} & \multicolumn{1}{r}{24.509} \\ \hline
Conformer with WavLM reg. \& m.n. & \multicolumn{1}{c|}{94 kh} & \multicolumn{1}{c|}{3.744} & \multicolumn{1}{c|}{\textbf{3.824}} & \multicolumn{1}{c|}{\textbf{3.218}} & \multicolumn{1}{c|}{3.806} & \multicolumn{1}{c|}{\textbf{3.771}} & \multicolumn{1}{c|}{3.251} & \multicolumn{1}{c|}{\textbf{7.730}}    & \multicolumn{1}{r}{\textbf{20.720}} \\ \hlineB{3}
Conformer with WavLM reg. & \multicolumn{1}{c|}{960 h} & \multicolumn{1}{c|}{\textbf{3.748}} & \multicolumn{1}{c|}{3.795} & \multicolumn{1}{c|}{3.211} & \multicolumn{1}{c|}{\textbf{3.810}} & \multicolumn{1}{c|}{3.721} & \multicolumn{1}{c|}{3.247} & \multicolumn{1}{c|}{7.298}    & \multicolumn{1}{r}{25.829} \\ \hline 
Conformer with WavLM reg. \& m.n. & \multicolumn{1}{c|}{960 h} & \multicolumn{1}{c|}{3.745} & \multicolumn{1}{c|}{\textbf{3.844}} & \multicolumn{1}{c|}{\textbf{3.220}} & \multicolumn{1}{c|}{\textbf{3.810}} & \multicolumn{1}{c|}{\textbf{3.772}} & \multicolumn{1}{c|}{\textbf{3.254}} & \multicolumn{1}{c|}{\textbf{7.783}}    & \multicolumn{1}{r}{\textbf{20.020}} \\ \hline \hline
\multicolumn{9}{c}{\textbf{\textit{Limited amount of noise resource setup (10 $\%$ noise data)}}}\\ \hline \hline
LSTM       & \multicolumn{1}{c|}{-} & \multicolumn{1}{c|}{3.795} & \multicolumn{1}{c|}{3.891} & \multicolumn{1}{c|}{3.290} & \multicolumn{1}{c|}{3.844} & \multicolumn{1}{c|}{3.766} & \multicolumn{1}{c|}{3.257} & \multicolumn{1}{c|}{8.749}    & \multicolumn{1}{r}{20.900} \\ \hline
LSTM with WavLM        & \multicolumn{1}{c|}{94 kh} & \multicolumn{1}{c|}{3.811} & \multicolumn{1}{c|}{3.915} & \multicolumn{1}{c|}{3.303} & \multicolumn{1}{c|}{3.906} & \multicolumn{1}{c|}{3.880} & \multicolumn{1}{c|}{3.368} & \multicolumn{1}{c|}{8.820}    & \multicolumn{1}{r}{20.670} \\ \hline
LSTM with WavLM m.n. & \multicolumn{1}{c|}{94 kh} & \multicolumn{1}{c|}{3.807} & \multicolumn{1}{c|}{3.929} & \multicolumn{1}{c|}{3.308} & \multicolumn{1}{c|}{3.908} & \multicolumn{1}{c|}{3.893} & \multicolumn{1}{c|}{3.368} & \multicolumn{1}{c|}{8.908}    & \multicolumn{1}{r}{20.035} \\ \hline
LSTM with WavLM reg. & \multicolumn{1}{c|}{94 kh} & \multicolumn{1}{c|}{3.803} & \multicolumn{1}{c|}{3.922} & \multicolumn{1}{c|}{3.314} & \multicolumn{1}{c|}{3.912} & \multicolumn{1}{c|}{3.911} & \multicolumn{1}{c|}{3.397} & \multicolumn{1}{c|}{8.906}    & \multicolumn{1}{r}{20.517} \\ \hline
LSTM with WavLM reg. \& m.n. & \multicolumn{1}{c|}{94 kh} & \multicolumn{1}{c|}{\textbf{3.818}} & \multicolumn{1}{c|}{\textbf{3.938}} & \multicolumn{1}{c|}{\textbf{3.325}} & \multicolumn{1}{c|}{\textbf{3.936}} & \multicolumn{1}{c|}{\textbf{3.932}} & \multicolumn{1}{c|}{\textbf{3.411}} & \multicolumn{1}{c|}{\textbf{9.105}}    & \multicolumn{1}{r}{\textbf{17.610}} \\ \hlineB{3}
LSTM with WavLM reg.       & \multicolumn{1}{c|}{960 h} & \multicolumn{1}{c|}{\textbf{3.833}} & \multicolumn{1}{c|}{3.921} & \multicolumn{1}{c|}{3.324} & \multicolumn{1}{c|}{3.937} & \multicolumn{1}{c|}{3.921} & \multicolumn{1}{c|}{3.403} & \multicolumn{1}{c|}{8.996}    & \multicolumn{1}{r}{20.141} \\ \hline
LSTM with WavLM reg. \& m.n. & \multicolumn{1}{c|}{960 h} & \multicolumn{1}{c|}{3.828} & \multicolumn{1}{c|}{\textbf{3.947}} & \multicolumn{1}{c|}{\textbf{3.326}} & \multicolumn{1}{c|}{\textbf{3.944}} & \multicolumn{1}{c|}{\textbf{3.959}} & \multicolumn{1}{c|}{\textbf{3.418}} & \multicolumn{1}{c|}{\textbf{9.246}}    & \multicolumn{1}{r}{\textbf{16.810}} \\ \hlineB{3}
Conformer  & \multicolumn{1}{c|}{-} & \multicolumn{1}{c|}{3.841} & \multicolumn{1}{c|}{3.918} & \multicolumn{1}{c|}{3.323} & \multicolumn{1}{c|}{3.982} & \multicolumn{1}{c|}{3.937} & \multicolumn{1}{c|}{3.446} & \multicolumn{1}{c|}{8.954}    & \multicolumn{1}{r}{19.006} \\ \hline
Conformer with WavLM   & \multicolumn{1}{c|}{94 kh} & \multicolumn{1}{c|}{3.841} & \multicolumn{1}{c|}{3.922} & \multicolumn{1}{c|}{3.333} & \multicolumn{1}{c|}{3.984} & \multicolumn{1}{c|}{3.941} & \multicolumn{1}{c|}{3.455} & \multicolumn{1}{c|}{9.066}    & \multicolumn{1}{r}{18.692} \\ \hline
Conformer with WavLM m.n. & \multicolumn{1}{c|}{94 kh} & \multicolumn{1}{c|}{3.845} & \multicolumn{1}{c|}{3.933} & \multicolumn{1}{c|}{3.343} & \multicolumn{1}{c|}{\textbf{3.989}} & \multicolumn{1}{c|}{3.942} & \multicolumn{1}{c|}{\textbf{3.458}} & \multicolumn{1}{c|}{9.079}    & \multicolumn{1}{r}{18.552} \\ \hline
Conformer with WavLM reg.  & \multicolumn{1}{c|}{94 kh} & \multicolumn{1}{c|}{3.842} & \multicolumn{1}{c|}{3.919} & \multicolumn{1}{c|}{3.335} & \multicolumn{1}{c|}{3.944} & \multicolumn{1}{c|}{3.930} & \multicolumn{1}{c|}{3.440} & \multicolumn{1}{c|}{8.997}    & \multicolumn{1}{r}{18.543} \\ \hline
Conformer with WavLM reg. \& m.n. & \multicolumn{1}{c|}{94 kh} & \multicolumn{1}{c|}{\textbf{3.861}} & \multicolumn{1}{c|}{\textbf{3.960}} & \multicolumn{1}{c|}{\textbf{3.345}} & \multicolumn{1}{c|}{3.978} & \multicolumn{1}{c|}{\textbf{3.952}} & \multicolumn{1}{c|}{3.456} & \multicolumn{1}{c|}{\textbf{9.184}}    & \multicolumn{1}{r}{\textbf{16.869}} \\ \hlineB{3}
Conformer with WavLM reg.& \multicolumn{1}{c|}{960 h}& \multicolumn{1}{c|}{\textbf{3.842}} & \multicolumn{1}{c|}{3.919} & \multicolumn{1}{c|}{3.335} & \multicolumn{1}{c|}{\textbf{3.981}} & \multicolumn{1}{c|}{3.931} & \multicolumn{1}{c|}{3.444} & \multicolumn{1}{c|}{9.003}    & \multicolumn{1}{r}{18.709} \\ \hline  
Conformer with WavLM reg. \& m.n. & \multicolumn{1}{c|}{960 h} & \multicolumn{1}{c|}{3.832} & \multicolumn{1}{c|}{\textbf{3.958}} & \multicolumn{1}{c|}{\textbf{3.343}} & \multicolumn{1}{c|}{3.980} & \multicolumn{1}{c|}{\textbf{3.955}} & \multicolumn{1}{c|}{\textbf{3.456}} & \multicolumn{1}{c|}{\textbf{9.192}}    & \multicolumn{1}{r}{\textbf{16.341}} \\ \hline \hline
\end{tabular}
\label{tab:low}
\end{table*}

\begin{table*}[t] \small
\centering
\caption{Results on the high fine-tuning resource setup with WavLM and its variants pre-trained with 94k data. The reg. and m.n denote the regression loss and noise mixing training strategy, respectively.}
\begin{tabular}{l|lll|lllll}
\hline \hline
\multicolumn{1}{c|}{\multirow{2}{*}{Model}} & \multicolumn{3}{c|}{DNS3 test data} & \multicolumn{5}{c}{Simulated test data} \\ \cline{2-9} 
 & \multicolumn{1}{c|}{SIG} & \multicolumn{1}{c|}{BAK} & \multicolumn{1}{c|}{OVR} & \multicolumn{1}{c|}{SIG} & \multicolumn{1}{c|}{BAK} & \multicolumn{1}{c|}{OVR} & \multicolumn{1}{c|}{SDR} & WER \\ \hline

Noisy      & \multicolumn{1}{c|}{3.776} & \multicolumn{1}{c|}{3.208} & \multicolumn{1}{c|}{3.114} & \multicolumn{1}{c|}{3.837} & \multicolumn{1}{c|}{2.976} & \multicolumn{1}{c|}{3.098} & \multicolumn{1}{c|}{3.598} & \multicolumn{1}{r}{24.721}  \\ \hline
Oracle magnitude mask  & \multicolumn{1}{c|}{-} & \multicolumn{1}{c|}{-} & \multicolumn{1}{c|}{-} & \multicolumn{1}{c|}{3.918} & \multicolumn{1}{c|}{4.151} & \multicolumn{1}{c|}{3.539} & \multicolumn{1}{c|}{12.537}    &  \multicolumn{1}{r}{2.596}   \\ \hlineB{3}
LSTM       & \multicolumn{1}{c|}{3.834} & \multicolumn{1}{c|}{4.079} & \multicolumn{1}{c|}{3.414} &  \multicolumn{1}{c|}{3.933} & \multicolumn{1}{c|}{4.009} & \multicolumn{1}{c|}{3.412} & \multicolumn{1}{c|}{9.630}    & \multicolumn{1}{r}{19.198} \\ \hline
LSTM with WavLM        & \multicolumn{1}{c|}{3.847} & \multicolumn{1}{c|}{4.073} & \multicolumn{1}{c|}{3.416} & \multicolumn{1}{c|}{3.941} & \multicolumn{1}{c|}{4.012} & \multicolumn{1}{c|}{3.418} & \multicolumn{1}{c|}{9.753}    & \multicolumn{1}{r}{18.008} \\ \hline
LSTM with WavLM m.n. & \multicolumn{1}{c|}{3.842} & \multicolumn{1}{c|}{4.081} & \multicolumn{1}{c|}{3.417} & \multicolumn{1}{c|}{3.937} & \multicolumn{1}{c|}{4.017} & \multicolumn{1}{c|}{3.419} & \multicolumn{1}{c|}{9.757}    & \multicolumn{1}{r}{18.017} \\ \hline
LSTM with WavLM reg.       & \multicolumn{1}{c|}{3.843} & \multicolumn{1}{c|}{4.073} & \multicolumn{1}{c|}{3.416} & \multicolumn{1}{c|}{3.935} & \multicolumn{1}{c|}{4.005} & \multicolumn{1}{c|}{3.415} & \multicolumn{1}{c|}{9.728}    & \multicolumn{1}{r}{18.251} \\ \hline
LSTM with WavLM reg. \& m.n. & \multicolumn{1}{c|}{\textbf{3.848}} & \multicolumn{1}{c|}{\textbf{4.094}} & \multicolumn{1}{c|}{\textbf{3.424}} & \multicolumn{1}{c|}{\textbf{3.942}} & \multicolumn{1}{c|}{\textbf{4.024}} & \multicolumn{1}{c|}{\textbf{3.426}} & \multicolumn{1}{c|}{\textbf{9.800}}    & \multicolumn{1}{r}{\textbf{17.521}} \\ \hlineB{3}
Conformer  & \multicolumn{1}{c|}{3.850} & \multicolumn{1}{c|}{4.090} & \multicolumn{1}{c|}{3.437} &  \multicolumn{1}{c|}{3.978} & \multicolumn{1}{c|}{\textbf{4.079}} & \multicolumn{1}{c|}{3.493} & \multicolumn{1}{c|}{9.714}    & \multicolumn{1}{r}{16.641} \\ \hline
Conformer with WavLM   & \multicolumn{1}{c|}{3.862} & \multicolumn{1}{c|}{4.092} & \multicolumn{1}{c|}{3.442} & \multicolumn{1}{c|}{3.982} & \multicolumn{1}{c|}{4.063} & \multicolumn{1}{c|}{3.496} & \multicolumn{1}{c|}{9.820}    & \multicolumn{1}{r}{16.104} \\ \hline
Conformer with WavLM m.n. & \multicolumn{1}{c|}{3.862} & \multicolumn{1}{c|}{4.091} & \multicolumn{1}{c|}{3.446} & \multicolumn{1}{c|}{3.981} & \multicolumn{1}{c|}{4.056} & \multicolumn{1}{c|}{3.491} & \multicolumn{1}{c|}{9.812}    & \multicolumn{1}{r}{16.279} \\ \hline
Conformer with WavLM reg.  & \multicolumn{1}{c|}{\textbf{3.863}} & \multicolumn{1}{c|}{4.097} & \multicolumn{1}{c|}{3.460} & \multicolumn{1}{c|}{\textbf{3.984}} & \multicolumn{1}{c|}{4.071} & \multicolumn{1}{c|}{3.497} & \multicolumn{1}{c|}{9.839}    & \multicolumn{1}{r}{16.087} \\ \hline 
Conformer with WavLM reg. \& m.n. & \multicolumn{1}{c|}{3.860} & \multicolumn{1}{c|}{\textbf{4.099}} & \multicolumn{1}{c|}{\textbf{3.450}} & \multicolumn{1}{c|}{3.979} & \multicolumn{1}{c|}{4.062} & \multicolumn{1}{c|}{\textbf{3.498}} & \multicolumn{1}{c|}{\textbf{9.884}}    & \multicolumn{1}{r}{\textbf{15.982}} \\ \hline \hline
\end{tabular}
\label{tab:high}
\end{table*}

\section{Experiment}
\subsection{Experiment Setup}
We used the DNS challenge 3 blind test data (DNS3)~\cite{reddy21_interspeech} and a simulated corpus to evaluate enhancement performance. The simulated test set consisted of 60 hours of simulated audio with SNR ranging from $-10$ dB to $30$ dB, taken from a clean speech signal of the LibriSpeech test set and synthesized with an RIR. The simulated test data mixed the clean speech with both Gaussian noise and non-stationary noise. The non-stationary noise was generated by combining noise recordings of SoundBible and Freesound~\cite{fonseca2017freesound} convolved with RIRs.

To see the impact of the WavLM on speech enhancement, we use the Deep Noise Suppression Mean Opinion Score (DNSMOS) P.835~\cite{reddy2022dnsmos} as an evaluation metric for both test data. Additionally, we evaluate the signal-to-distortion ratio (SDR) and WER for the simulated test data as the transcriptions and clean references are available. 
In the DNSMOS P.835, there are three output scores: i) speech quality (SIG), ii) background noise quality (BAK), and iii) overall quality (OVR). A higher score indicates better enhancement quality. For the ASR evaluation, 
we fed the enhanced speech to a pre-trained HuBERT-based ASR inference model~\cite{hsu2021hubert} with a 4-gram language model to calculate the WER. 
The hyperparameters and setting of the ASR inference model are identical with~\cite{baevski2020wav2vec2}.

\subsection{Evaluation Results for Low Fine-Tuning Resource}

Table~\ref{tab:low} shows the speech enhancement results using the WavLM pre-trained with 94k hours in the low-resource setting, where reg. and m.n. denote the regression loss and noise mixing strategy in pre-training, respectively.
In the limited amount of speech resource setup (10 hours), we can observe WavLM helped substantially improve both the WER and speech quality compared to the baseline. If we only apply either the noise mixing or the regression during pre-training, the WER was further reduced while there was no significant gain on the speech quality. The best result was obtained by applying both the noise mixing and regression loss, achieving a 0.1 DNSMOS OVR gain, a 1.4 SDR improvement, and a relative 24.8$\%$ WER reduction for the LSTM task-specific layers, indicating the effectiveness of SSL for the low-resource speech enhancement scenario. 

The results for the limited amount of noise setup (10\% noise) are also shown in the table. The best result achieved a 0.1 DNSMOS OVR gain, a 0.35 SDR improvement, and a relative 15.7$\%$ WER reduction with the LSTM enhancement model. The overall performance improvements were smaller than those of the previous experiment (i.e., the 10-hour setup). This means that SSL can better compensate for the lack of sufficient speech data for the enhancement. Nonetheless, it also provides modest improvement when the noise training data are scarce.

\subsection{Evaluation Results for High Fine-Tuning Resource}
Table~\ref{tab:high} shows the results for the high fine-tuning resource setting.
The best result achieved a 0.01 DNSMOS OVR gain, a 0.17 SDR improvement, and a relative 8.7$\%$ WER reduction for the LSTM enhancement model.
In terms of DNSMOS, the speech enhancement model with SSL appears to have approached the oracle mask performance, which may explain the limited improvement. On the other hand, there was still a large WER gap between the oracle mask and the enhancement models, and the pre-training helped reduce this gap significantly. 

In the high fine-tuning resource setting, WavLM variant with the noise mixing strategy is better than the classification-based loss, in terms of WER reduction and speech quality, which is consistent with the observation in the low-resource setting. Unlike other speech downstream tasks, the regression-based self-supervised learning loss is the best one. A possible explanation is the regression-based loss with denoising pre-training task is similar to the enhancement task, so the model will learn related information.

\subsection{Impact of Different Amounts of Pre-Training Data}

Finally, we examine the impact of the quantity of the WavLM pre-training data. For the regression-based WavLM with and without noise mixing, we show the speech enhancement results based on the WavLM models trained on the LibriSpeech 960 hours data in Table 1. By comparing these with the corresponding results obtained with the WavLM models trained on the 94k-hour data, we can see that they produced similar speech enhancement results despite the 100x scale difference in the pre-training data. We argue that the lack of the performance improvement from using a larger pre-training dataset can be attributed to the fact that Libri-Light~\cite{kahn2020libri}, which accounts for a large portion (60k hours) of the 94-hours data, consists of noisy speech data with SNRs ranging approximately between 0 dB and 20 dB. During the pre-training, these noisy speech data were included as the clean prediction target. This may have resulted in the lack of the speech enhancement performance improvement. This phenomenon raises a limitation of the current pre-training approach using a huge amount of noisy data, calling for the development of more tailored pre-training methods. 

\section{Conclusions}

In this paper, we explored the effect of WavLM on speech enhancement. We proposed two WavLM modifications: using the regression-based training objective and the noise mixing strategy during pre-training.
In the fine-tuning stage, we used WavLM as a feature extractor and applied speech enhancement models as the task-specific layers, where we considered two enhancement models to obtain generalizable insights.
A series of experiments were carried out to explore the efficacy of WavLM and its variants under three conditions about the amount of data.
The fine-tuned models were evaluated on DNS3 and the simulated test data in terms of both speech quality and WER.
It was shown that WavLM provided substantial speech enhancement improvement in terms of the WER while the speech quality metric improvement was rather modest.
The potential limitation of the proposed pre-training approach was also discussed, calling for further investigation to develop pre-training schemes that are optimal for the speech enhancement tasks~\cite{wang2020self,xiang2020CCGANSE,wisdom2020unsupervised}. 

\bibliographystyle{IEEEtran}

\bibliography{main}


\end{document}